\documentclass[preprint,12pt,3p]{elsarticle}

\usepackage{matlab-prettifier}
\usepackage{graphicx}
\usepackage{makecell}
\usepackage{xfrac}
\usepackage{parskip}
\usepackage{adjustbox}
\usepackage{setspace}
\usepackage{hyperref}
\hypersetup{ 
pdftoolbar=true, 
pdfmenubar=true, 
pdffitwindow=false, 
pdfstartview={FitH}, 
pdftitle={}, 
pdfauthor={}, 
pdfsubject={}, 
pdfkeywords={}{}, 
pdfnewwindow=true, 
colorlinks=true, 
linkcolor=BlueViolet, 
citecolor=maroon, 
urlcolor=blue,
}
\usepackage{comment}
\usepackage{subfig}
\usepackage{caption}
\usepackage{longtable}
\usepackage{amssymb}
\usepackage{amsmath}
\usepackage{gensymb}
\usepackage{siunitx}
\usepackage{CJKutf8}         

\usepackage{natbib}
\biboptions{comma,round}
\setcitestyle{authoryear}

\begin{document}
\begin{frontmatter}

\title{Towards a representative social cost of carbon}

\author[label0]{Jinchi Dong}
\author[label1,label2,label3,label4,label5,label6,label7]{Richard S.J. Tol\corref{cor1}
}
\author[label8,label9]{Fangzhi Wang}\fnref{label10}
\address[label0]{School of the Environment, Nanjing University, China}
\address[label1]{Department of Economics, University of Sussex, Falmer, United Kingdom}
\address[label2]{Institute for Environmental Studies, Vrije Universiteit, Amsterdam, The Netherlands}
\address[label3]{Department of Spatial Economics, Vrije Universiteit, Amsterdam, The Netherlands}
\address[label4]{Tinbergen Institute, Amsterdam, The Netherlands}
\address[label5]{CESifo, Munich, Germany}
\address[label6]{Payne Institute for Public Policy, Colorado School of Mines, Golden, CO, USA}
\address[label7]{College of Business, Abu Dhabi University, UAE}
\address[label8]{School of Management and Economics, Beijing Institute of Technology, Beijing, 100081, China}
\address[label9]{Center for Energy and Environmental Policy Research, Beijing Institute of Technology, Beijing, 100081, China}

\cortext[cor1]{Jubilee Building, BN1 9SL, UK}
\fntext[label10]{We are grateful to Moritz Drupp for sharing unpublished data from the expert survey.}

\ead{r.tol@sussex.ac.uk}
\ead[url]{http://www.ae-info.org/ae/Member/Tol\_Richard}

\begin{abstract}
The majority of estimates of the social cost of carbon use preference parameters calibrated to data for North America and Europe. We here use representative data for attitudes to time and risk across the world. The social cost of carbon is substantially higher in the global north than in the south. The difference is more pronounced if we count people rather than countries.
\end{abstract}

\end{frontmatter}
\newpage \section{Introduction}
The social cost of carbon is the damage done, at the margin, by the emission of carbon dioxide. Many assumptions are used to estimate the social cost of carbon. Most of these assumptions are positive; the climate sensitivity is a prime example. Some assumptions are normative; the pure rate of time preference comes to mind. Reasonable people can reasonably disagree about the social welfare function \citep[indeed,][shows they cannot agree]{Arrow1950}. An individual's ethical views are partly idiosyncratic and partly cultural. Norms about time and risk have been found to systematically vary between countries. The social cost of carbon, however, is primarily estimated by researchers from North America and Western Europe. In this paper, we recalibrate the social cost of carbon according to the stated preferences of people from 76 countries across the world.

\begin{figure}
    \centering
    \caption{Share of papers published on the social cost of carbon by country of affiliation.}
    \includegraphics[width=\textwidth]{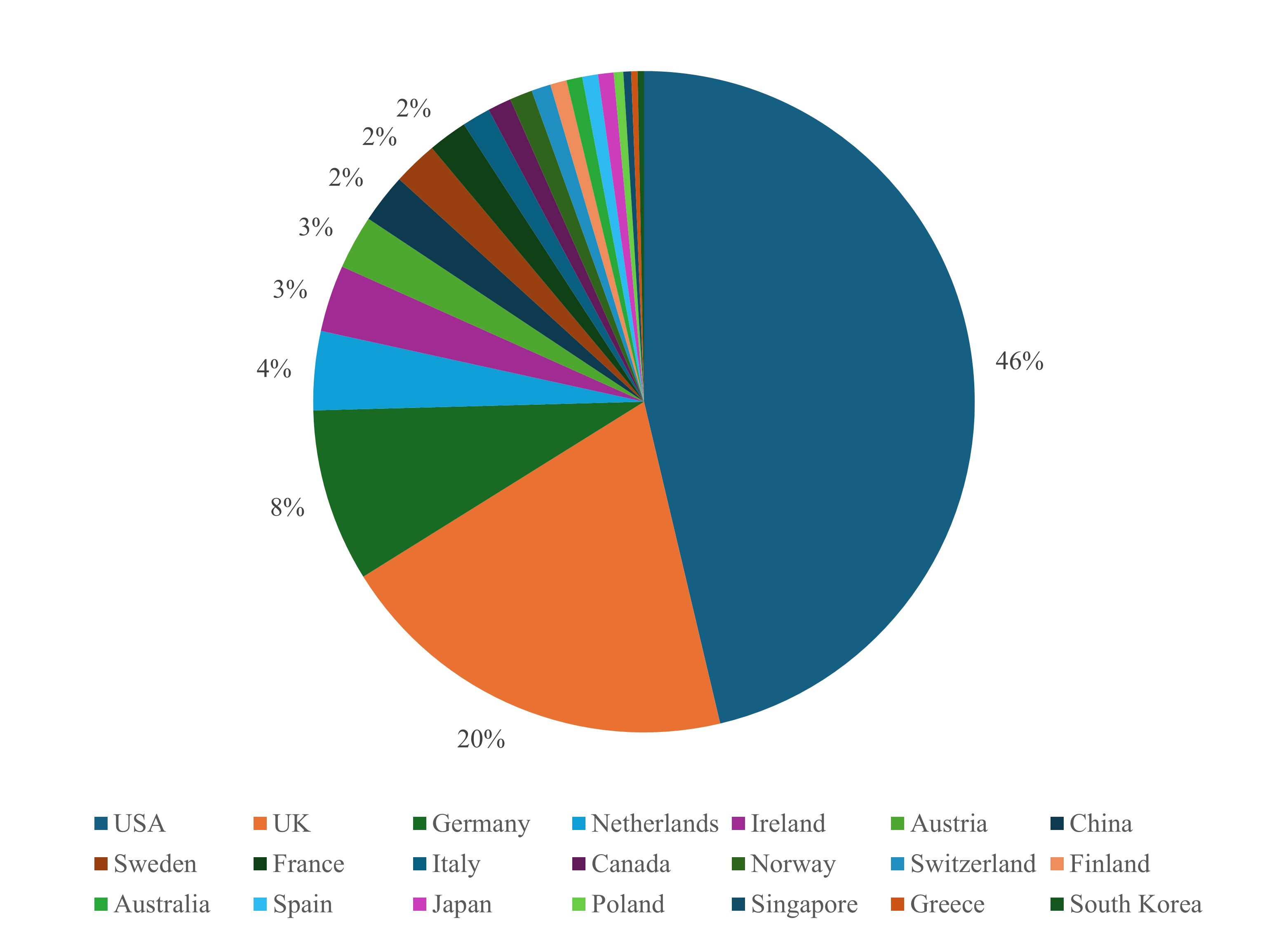}
    \caption*{\footnotesize Papers published between 1980 and 2023. Papers are attributed to country of affiliation at the time of publication and inversely proportional to the number of co-authors. Source: \citet{Tol2024data}.}
    \label{fig:meta}
\end{figure}

Figure \ref{fig:meta} groups 323 papers on the social cost of carbon by the country of affiliation of the authors of these papers \citep[data from][]{Tol2024data}. Papers of mixed nationality are attributed proportionally to the number of authors. The USA contributed most (46\%) followed by the UK (20\%). Africa and Latin America did not contribute to this literature. Only three non-Western countries are represented, all from East Asia. There is no reason to believe that people from different parts of the world would systematically differ in their interpretation of the evidence about climate change and its impact, but they may well hold different attitudes to the future. The literature on the social cost of carbon may thus be biased towards Western attitudes.

``Western attitudes'' contain a multitude, of course. There has been a lively debate on the pure rate of time preference in the context of climate policy, first between \citet{Nordhaus1992} and \citet{Cline1992}, and later between \citet{Stern2006} and \citet{Nordhaus2007jel}. \citet{Arrow1996IPCC} described this as a choice between descriptive and prescriptive discounting. The range of opinions expressed in \citet{Drupp2018} shows that the debate has not abated. In fact, the discussion has become more complicated as alternatives to exponential discounting and its measurement have emerged \citep{Cropper1991, Weitzman2001, Newell2003, Tol2013EL, Giglio2015, Iverson2021, Jaakkola2022, Bauer2023, Eden2023}.

Participants in this debate draw, almost exclusively, from Western cultures. The \emph{prescriptive} school relies, essentially, on Aristotle's verdict against usury, which was later adopted by St Augustine and the Prophet Muhammad. The \emph{descriptive} school typically calibrates time and risk preferences with data for the market for U.S. Treasuries. \citet{Sohn2019} is an exception, estimating the social cost of carbon using time and risk preferences based on data from South Korea. In an attempt to reflect a wider range of opinions, \citet{Anthoff2009erl} reflect a wider range of opinions, using the 20 OECD countries in \citet{Evans2005FS}. Data for the rest of the world has since improved considerably and this allows us here to cast a wider net and so be more inclusive and representative of the world population.

The paper proceeds as follows. Section \ref{sc:method} presents the data and methods. Section \ref{sc:results} discusses the results. Section \ref{sc:conclude} concludes.

\section{Materials and methods}
\label{sc:method}

\subsection{Data and calibration}
\citet{Falk2018} and \citet{Falk2023} report attitudes towards time and risk for 76 countries,\footnote{The data are \href{https://www.briq-institute.org/global-preferences/home}{here}.} which together constitute 85\% of the world population and 93\% of the world economy. These preferences are stated in responses to intuitive questions in an unincentivized survey. Falk \textit{et al.} show that these simple measures correlate well with the results of state-of-the-art preference elicitation in surveys and experiments. They further show that stated preferences do not systematically differ from revealed preferences in incentivized elicitation. Figure \ref{fig:falk} shows the indicators of patience and risk-taking, aggregated to the country level. The two measures are largely uncorrelated, with the possible exception for extreme impatience and risk aversion.

Note that, Falk \textit{et al.} report indices for, rather than rates of, time preference and risk aversion. \citet{Sunde2022} calibrate rates to indices. We calibrate Falk's data to the results of \citet{Drupp2018}, who surveyed 181 economists about the appropriate pure rate of time preference and the elasticity of intertemporal substitution, the two parameters in the \citet{Ramsey1928} rule of discount. There are surveys of time preference \citep{Frederick2002, Wang2016JEP} and the elasticity of marginal utility \citep{Havranek2015JIE, Havranek2015JEEA, Thimme2017}, but Drupp's is the only study we know that considers both together. We use a linear calibration, matching the 5\% and 95\%ile so that 90\% of the imputed data are \emph{interpolated} and only 10\% \emph{extrapolated}.

No calibration is perfect. The appendix details five alternative calibrations. One restricts the sample to Europe and North America, the location of most of Drupp's experts. A second alternative weighs Falk's country data by population size. The remaining three calibrations use alternatives to Drupp's data.

\subsection{Model}
We use \textsc{fund4.0m-g}, one of a stable of integrated assessment models collectively known as \textsc{fund}. This is a global model implemented in \textsc{Matlab}. The model consists of a \citet{Solow1956} growth model with energy (and associated emissions of carbon dioxide) as a derived demand rather than a factor of production, as in \textsc{dice} \citep{Nordhaus1993}. Emissions are input into a \citet{MaierReimer1987} carbon cycle model which is coupled to a \citet{Schneider1981} climate model. Climate change feeds back on total factor productivity according to a damage function calibrated as in \citet{Barrage2023}. The main departure from Nordhaus' seminal work is that we assume that society becomes less vulnerable to climate change as incomes grow \citep{Schelling1984}. This version of the model has been used previously \citep{Tol2020handbook, Tol2024EnPol}; earlier versions date back to \citet{Tol1997}.

The social cost of carbon is computed as the difference in relative impacts due to a small increase in carbon dioxide emissions in 2015, multiplied by total output in the unperturbed scenario, discounted with the Ramsey rate.

\section{Results}
\label{sc:results}
\begin{figure}
    \centering
    \caption{Cumulative frequency of the global social cost of carbon with national time and risk preferences.}
    \includegraphics[width=\textwidth]{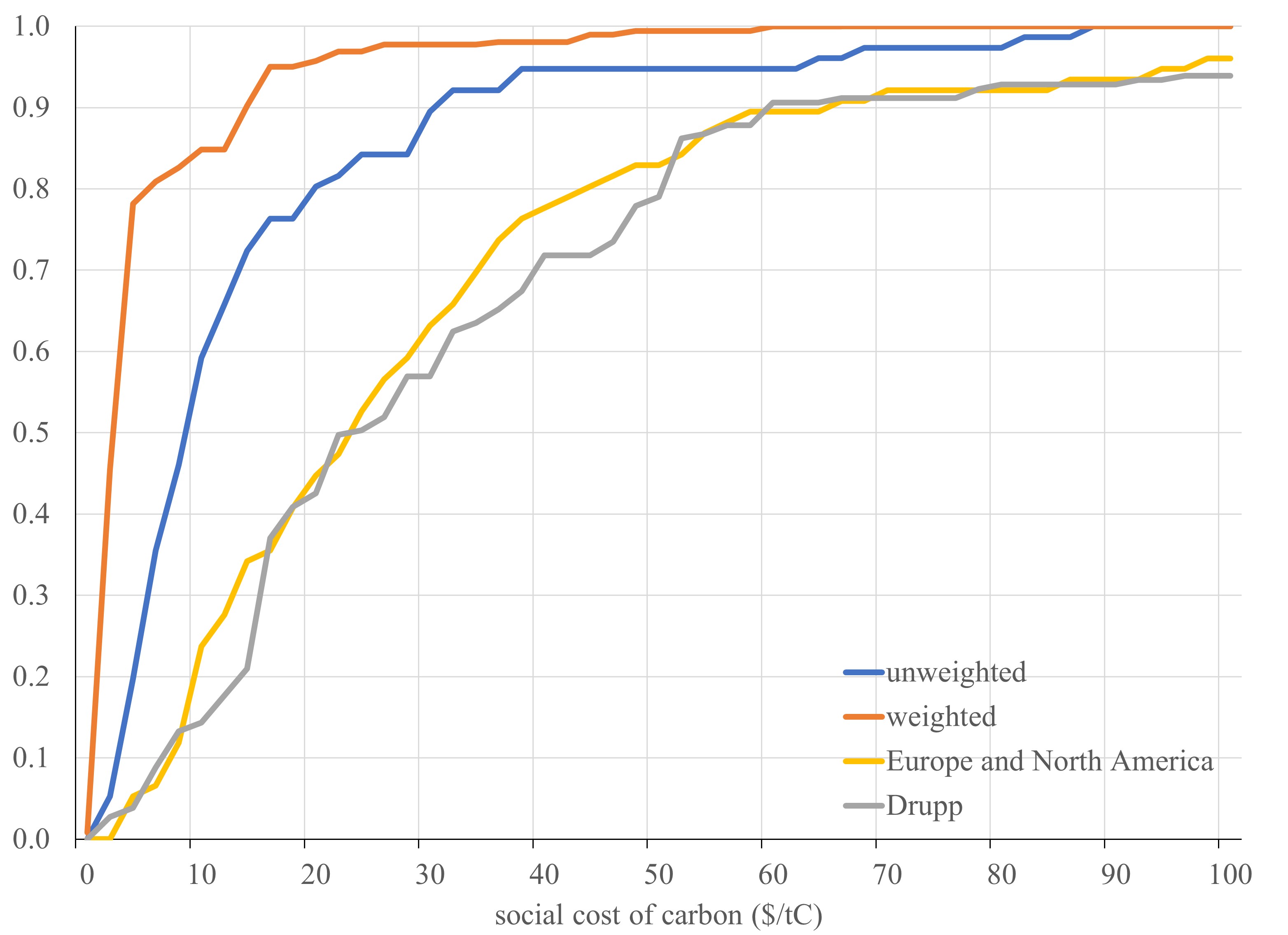}
    \caption*{\footnotesize Three cases are shown: The calibration of the Falk data to the Drupp data, the population-weighted calibration of the Falk data to the population-weighted Drupp data, and Drupp's expert data.}
    \label{fig:cdf}
\end{figure}

Figure \ref{fig:cdf} shows the cumulative frequency of the social cost of carbon for all 76 countries of the Falk data for the SSP2 scenario, which is the most likely scenario according to \citet{Srikrishnan2022}. These estimates reflect what the global social planner would do if she assumed the average preferences of the people in one of the listed countries. The median country prefers a carbon tax of about \$10/tC, with opinions ranging from \$2/tC (5th \%ile) to \$60/tC (95th \%ile).

For comparison, Figure \ref{fig:cdf} includes the cumulative frequency for Drupp's 181 experts. The median scholar favours \$24/tC, much higher than the median country. Geographically limiting the calibration sample makes the global distribution look much like Drupp's sample. Population-weighting shifts the entire distribution to the left. The median person would prefer a carbon tax of \$4/tC. 

The median scholar is in line, unsurprisingly, with the scholarly literature. \citet{Stern2006} advocate $\rho=0.001$ and $\eta=1$, which yields a social cost of carbon of \$48.8/tC. \citet{OMB2023} sets the money discount rate at 2\%, which \citet{Nyserda2021} argue implies $\rho=0.003$ and $\eta=1.3$. The social cost of carbon is then \$30.9/tC. \citet{Nordhaus1992} prefers $\rho=0.03$ and $\eta=1$, resulting in \$15.4/tC while \citet{Nordhaus2014} favours $\rho=0.015$ and $\eta=1.5$ and a social cost of carbon of \$15.8/tC. \citet{Weitzman2007} is most in line with popular opinion, arguing for $\rho= 0.02$ and $\eta=2$ and so \$8.8/tC.

Averaged over Drupp's experts, the social cost of carbon is \$34.8/tC. Averaged over the countries, the social cost of carbon is much lower, \$15.4/tC. It is higher, however, than the population-weighted average, \$12.2/tC. This is because more populous nations tend to be more impatient. If we use average preferences, the social cost of carbon falls further, to \$9.3/tC because the social cost of carbon increases disproportionally with decreasing time and risk preferences.

\section{Discussion and conclusion}
\label{sc:conclude}
The literature on the social cost of carbon is dominated by Western scholars. We calibrate the pure rate of time preference and the rate of risk aversion to representative data for 76 countries. We find that Western scholars advocate a higher social cost of carbon than would national representatives. The social cost of carbon may fall further if we weigh the results by the number of people.

Having extensively surveyed the literature on the social cost of carbon \citep{Tol2023NCC, Tol2024data}, we are fairly sure that this is the first paper to try and establish a \emph{representative} social cost of carbon. It should not be the last. The Falk data are indicators of individual impatience and risk-taking that we \emph{assumed} to be proportional to the presumably social pure rate of time preference and elasticity of intertemporal substitution according to Drupp. These studies are, in our opinion, the best available but not without flaws. Connecting two disparate datasets is never easy, but we here have a mismatch of both what is measured and where. In the appendix, we also find somewhat different results for the Hofstede data and the literature review. The analysis here should therefore be repeated when better data become available. We abstract from uncertainty and inequity, and so dodge the question whether the inverse of the rate of intertemporal substitution equals the rate of risk aversion and the pure rate of inequity aversion \citep{Agneman2024, Anthoff2019, Ha-Duong2004, Saelen2009, Tol2010}. Neither the Drupp nor the Falk data allow us to make this distinction. We explored only a small part of the parameter and model space of the social cost of carbon: We use a model with a single region and a single sector; we abstract from the impact of climate change on economic growth; we ignore uncertainty, ambiguity, and stochasticity; we omit fat tails and tipping points; and so on. More research is therefore needed but\textemdash since none of these extensions affects the relationship between the social cost of carbon on the one hand and the pure rate of time preference and the elasticity of intertemporal substitution on the other\textemdash we are confident that this would not detract from our key finding: The ethical values assumed by experts systematically deviate from the world population so that published estimates of the social cost of carbon are unrepresentative and too high.

\newpage \bibliography{master}
\newpage \setcounter{page}{1}
\renewcommand{\thepage}{A\arabic{page}} \appendix

\section{Calibration}
We impose the linear relationship $\rho_c = \gamma_\rho + \lambda_\rho r_c$ and $\eta_c = \gamma_\eta + \lambda_\eta e_c$ where $\rho$ is the pure rate of time preference, $\eta$ is the Arrow-Pratt rate of relative risk aversion, $r$ is Falk's index of patience, $e$ is Falk's index of risk-taking, $\gamma_i$ and $\lambda_i$ are calibration parameters, and $c$ denotes country.

In the central calibration, we choose $\gamma_i$ and $\lambda_i$ such that the $\rho_{(c)} = r_{(c)}$ and $\eta_{(c)} = e_{(c)}$ for $(c) = 0.05$ and $(c) = 0.95$ of the Drupp and Falk data. We then use the calibrated parameters to derive the welfare parameters for each country. See Table \ref{tab:calib}. Note that we impose the restrictions $\rho_c \geq 0$ and $\eta_c \geq 0$. We use the tails to calibrate so that 90\% of the imputed data are \emph{interpolated} and only 10\% \emph{extrapolated}.

Drupp's data are unrepresentative: 44\% of the surveyed experts are in North America, 49\% in Europe, and 7\% in the rest of the world. The corresponding numbers for the world population in 2020 are 6\%, 8\% and 86\%. This matters. In Falk's data, the patience (risk-taking) score is 0.80 (0.12) in North America, 0.28 (-0.09) in Europe and 0.00 (-0.07) in the rest of the world; these are population-weighted averages. We therefore weigh Drupp's data so that sample and population totals align. We similarly weigh Falk's data so that percentiles are for the world population. We then match the 5th and 95th percentiles of the weighted Drupp data with the weighted Falk data and continue as above. As an alternative solution to the lack of representation of Drupp's data, we restrict their data to experts from Europe and North America and similarly only consider Falk's data for these countries \emph{in the calibration}.

As a robustness check, we use the 2015 version of the cultural dimension data by \citet{Hofstede2003}.\footnote{The data are \href{https://geerthofstede.com/research-and-vsm/dimension-data-matrix/}{here}.} Particularly, we use long-term orientation as a proxy for pure time preference and uncertainty avoidance as a proxy for risk aversion and hence intertemporal substitution. Calibration is as for Falk. Hofstede has fewer countries (63) than Falk (76). More importantly, Hofstede focuses on \emph{corporate} culture whereas Falk is about culture in general. Hofstede interviewed people who work for multinationals, whereas Falk interviewed residents. This is particularly problematic as relatively fewer people work in large corporations in poorer countries, and fewer still in foreign-owned ones.

As another robustness check, we survey the literature on the social discount rate for selected countries. We started with a search on \textsc{Scopus} for ``social discount rate'' and continued with checking the references of the papers found. The results are in Table \ref{tab:sdr}. We regressed the recommended pure rate of time preference on Falk's index of impatience, and the rate of risk aversion on Falk's index of risk-taking (with dummies for Iran and Russia). See Table \ref{tab:calib}. We use two variants of this. In the first variant, we use the observed rates as given in Table \ref{tab:sdr}, or the average if there are multiple observations for a country; for those countries without any observations, we use the regression model. In the second variant, we use the regression model for all countries.

The Drupp data are for the pure rate of time preference and the intertemporal rate of substitution. Although based on a survey, the respondents are experts who can be expected to understand both these concepts and the implications of their choice. The main disadvantage of the Drupp data is its lack of representativeness, not just because PhD economists are not like the general population but also geographically.

The literature on the appropriate social discount rate for various countries is more representative than Drupp's data but suffers other drawbacks. The pure rate of time preference is sometimes set by the authors based on convention. In other cases, it is based on the mortality rate. Although this approach has its supporters \cite[e.g][]{Addicott2020} and is reasonable for \emph{private} decisions, it is not uncontroversial as a guide to \emph{public} policy. The curvature of the utility function is measured in various ways\textemdash expenditure on necessary goods and redistribution through taxes and benefits being the most common ones\textemdash but not quite the elasticity of intertemporal substitution. However, together the alternative calibrations reveal key sensitivities and potential biases.

Table \ref{tab:assumptions} shows the details for each country. The first two calibrations see substantial variation between countries whereas all countries cluster near the average for the third calibration, particularly so for the variant in which all parameters are imputed. The Hofstede calibration uses a different set of countries and is therefore shown separately in Table \ref{tab:hofstede}. Differences between countries are again substantial.

\begin{figure}
    \centering
    \caption{Index of attitudes to time and risk according to \citet{Falk2018}, by country.}
    \includegraphics[width=\textwidth]{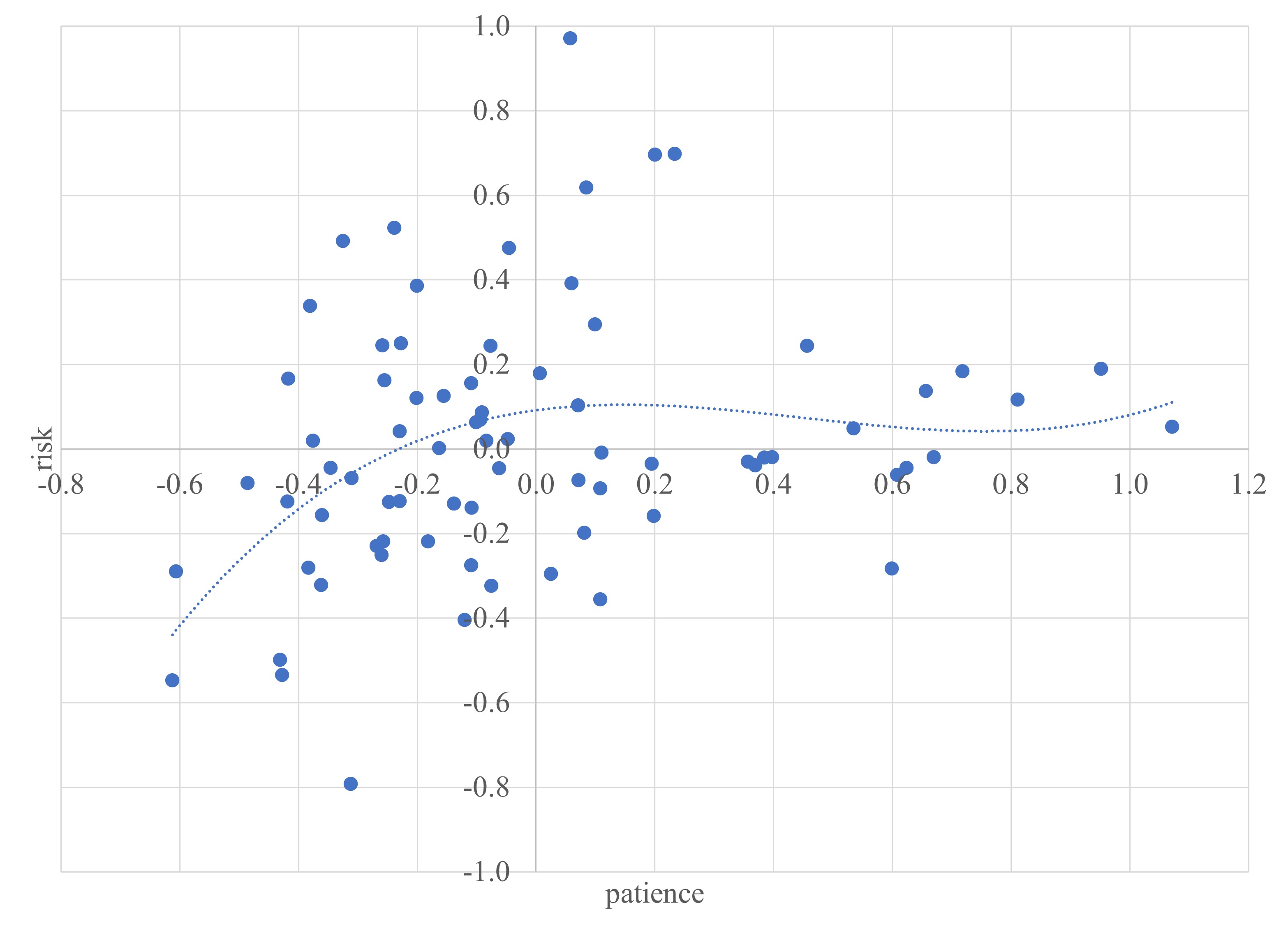}
    \label{fig:falk}
\end{figure}

\begin{table}
    \centering
    \caption{Published estimates of the social discount rate and its components for various countries}
    \tiny
    \begin{tabular}{l l c c c}
        Study & Country & $r$ & $\rho$ & $\eta$\\ \hline
        \citet{Moore2020sdr} & Argentina & 3.02 & 1 & 1.23 \\
        \citet{Villena2023} & Argentina & 7.20 & 1.21 & 1.41 \\
        \citet{Evans2005JES} & Austria & 5.3 & 1.0 & 1.6 \\
        \citet{Evans2004} & Australia & 4.7 & 1.5 & 1.4 - 1.7 \\
        \citet{Evans2005JES} & Belgium & 4.4 - 4.7 & 1.0 & 1.5 - 1.6 \\
        \citet{Moore2020sdr} & Bolivia & 2.81 & 1 & 1.06 \\
        \citet{Villena2023} & Bolivia & 3.70 & 1.23 & 1.25 \\
        \citet{Moore2020sdr} & Brazil & 4.36 & 1 & 2.09 \\
        \citet{Villena2023} & Brazil & 11.90 & 1.21 & 3.34 \\
        \citet{Boardman2010} & Canada & 3.5 & 1 & 1.5\\
        \citet{Moore2020sdr} & Chile & 5.72 & 1 & 1.22 \\
        \citet{Villena2023} & Chile & 6.50 & 1.23 & 1.33 \\
        \citet{Wang2013} & China & 4.5 & 0.672 & 1.07 \\
        \citet{Moore2020sdr} & Colombia & 4.61 & 1 & 1.61 \\
        \citet{Villena2023} & Colombia & 5.00 & 1.30 & 2.02 \\
        \citet{Moore2020sdr} & Costa Rica & 3.16 & 1 & 1.11 \\
        \citet{Evans2011} & Cyprus & 4.1 - 5.0 & 1 & 1.0 - 1.3 \\
        \citet{Evans2005JES} & Czech Republic & 3.1 & 1.1 & 1.4 \\
        \citet{Secilmis2019} & Czech Republic & 1.94 - 3.75 & 1.05 & 0.483 - 1.404\\
        \citet{Evans2005JES} & Denmark & 2.3 - 2.4 & 1.1 & 1.2 - 1.3\\
        \citet{Moore2020sdr} & Ecuador & 2.48 & 1 & 1.06 \\
        \citet{Villena2023} & Ecuador & 3.70 & 1.24 & 1.33 \\
        \citet{Moore2020sdr} & El Salvador & 3.43 & 1 & 1.13 \\
        \citet{Secilmis2019} & Estonia & 3.42 - 6.91 & 1.27 & 0.483 - 1.178\\
        \citet{Evans2005JES} & Finland & 4.5 & 1.0 & 1.6 \\
        \citet{Evans2004france, Evans2004, Evans2005JES} & France & 3.2 - 3.8 & 0.9 - 1.2 & 1.3 \\
        \citet{Evans2004, Evans2005JES} & Germany & 4.1 - 4.5 & 1.0 & 1.4 - 1.6 \\
        \citet{Schad2012} & Germany & 2.9 & 0.5 & 1.5 \\
        \citet{Evans2005JES} & Greece & 4.8 - 5.3 & 1.0 & 1.5 - 1.7 \\
        \citet{Moore2020sdr} & Guatemala & 2.72 & 1 & 1.04 \\
        \citet{Moore2020sdr} & Honduras & 3.35 & 1 & 1.16 \\
        \citet{Evans2005JES} & Hungary & 3.2 - 3.5 & 1.3 & 1.2 - 1.4\\
        \citet{Tabi2013} & Hungary & 4.1 & 1.27 & 1.4 \\
        \citet{Secilmis2019} & Hungary & 2.11 - 3.30 & 1.32 & 0.483 - 1.000\\
        \citet{Kula2004} & India & 5.2 & 1.3 & 1.64 \\
        \citet{Daneshmand2018} & Iran & 5.79 & 0.53 & 4.266 \\
        \citet{Evans2005JES} & Ireland & 5.6 - 6.8 & 0.8 & 1.6 - 2.0 \\
        \citet{Evans2005JES} & Italy & 4.5 - 4.7 & 1.0 & 1.4 - 1.5 \\
        \citet{Percoco2008} & Italy & 3.69 - 3.83 & 1 & 1.282 - 1.347 \\
        \citet{Evans2004} & Japan & 5.0 & 1.5 & 1.4 \\
        \citet{Secilmis2019} & Latvia & 3.50 - 6.67 & 1.42 & 0.483 - 1.092\\
        \citet{Kazlauskiene2004} & Lithuania & 3.75 & 1.37 & 0.53 \\
        \citet{Evans2005JES} & Luxembourg & 5.4 & 0.9 & 1.8 \\
        \citet{Moore2020sdr} & Mexico & 4.20 & 1 & 2.71 \\
        \citet{Evans2005JES} & Netherlands & 3.6 - 3.8 & 0.9 & 1.5 - 1.6 \\
        \citet{Moore2020sdr} & Nicaragua & 3.36 & 1 & 1.14 \\
        \citet{Moore2020sdr} & Panama & 3.61 & 1 & 1.15 \\
        \citet{Moore2020sdr} & Paraguay & 1.99 & 1 & 1.00 \\
        \citet{Villena2023} & Paraguay & 3.20 & 1.20 & 0.94 \\
        \citet{Moore2020sdr} & Peru & 3.93 & 1 & 1.05 \\
        \citet{Villena2023} & Peru & 6.30 & 1.26 & 1.72 \\
        \citet{Evans2005JES} & Poland & 6.1 & 1.0 & 1.1 \\
        \citet{Secilmis2019} & Poland & 2.75 - 4.94 & 0.98 & 0.483 - 1.085\\
        \citet{Foltyn-Zarychta2021} & Poland & 4.39 & 0.9599 & 1.1174 \\
        \citet{Evans2005JES} & Portugal & 5.3 - 5.6 & 1.0 & 1.6 - 1.7 \\
        \citet{Kossova2014, Kossova2016} & Russia & 3.2 - 3.9 & 1.48 & 0.2 \\
        \citet{Evans2005JES} & Slovakia & 6.6 - 7.0 & 1.1 & 1.5 - 1.6 \\
        \citet{Secilmis2019} & Slovakia & 2.53 - 5.32 & 0.98 & 0.483 - 1.353\\
        \citet{Sohn2019} & South Korea & 5.1 & 1.1 & 1.0 \\
        \citet{Evans2005JES} & Spain & 4.7 & 1.0 & 1.6 \\
        \citet{Evans2005JES} & Sweden & 2.4 - 2.8 & 1.1 & 1.1 - 1.4 \\
        \citet{Halicioglu2013} & Turkey & 5.06 & 0.61 & 1.686\\
        \citet{Akbulut2019} & Turkey & 4.41 - 4.88 & 0.58 & 1.0580 - 1.2042\\
        \citet{Evans2002, Evans2004, Evans2005JES} & United Kingdom & 4.2 - 4.8 & 1.0 & 1.5 - 1.6\\
        \citet{Groom2019} & United Kingdom & 4.5 & 1.5 & 1.5 \\
        \citet{Evans2004} & United States & 4.4 - 4.6 & 1.5 & 1.3 - 1.4 \\
        \citet{Moore2013} & United States & 3.5 & 1 & 1.35 \\
        \citet{Rennert2021} & United States & 3.0 & 0.8 & 1.5 \\
        \citet{Moore2020sdr} & Uruguay & 4.64 & 1 & 1.71 \\
        \citet{Villena2023} & Uruguay & 4.10 & 1.23 & 1.98 \\
        \citet{Moore2020sdr} & Venezuela & 2.81 & 1 & 1.04 \\ \hline
    \end{tabular}
    \caption*{\footnotesize The table shows the components of the Ramsey rule $r = \rho + \eta g$ where $r$ denotes the discount rate, $\rho$ the pure rate of time preference and $\eta$ the inverse of the elasticity of intertemporal substitution.}
    \label{tab:sdr}
\end{table}

\begin{table}
    \centering
    \caption{Indicators and rates of time and risk preferences and the corresponding calibrations.}
    \label{tab:calib}
    \begin{tabular}{l l c c c c | c c}
         & & $r$ & $e$ & $\rho$ & $\eta$ & constant & slope \\ \hline
         Falk & 5\%ile & -0.43 & -0.43 & 0 & 0.2 & $\gamma_\rho = 2.46$& $\lambda_\rho = 3.60$ \\
         &  95\%ile & 0.68 & 0.55 & 4 & 4 & $\gamma_\eta = 1.77$ & $\lambda_\eta = 2.87$\\
        Population weights & 5\%ile & -0.38 & -0.32 & 0 & 0.5 & $\gamma_\rho = 3.28$ & $\lambda_\rho = 4.57$\\
         & 95\%ile & 0.72 & 0.39 & 5 & 5 & $\gamma_\eta = 2.95$ & $\lambda_\eta = 6.34$\\
         N America \& Europe & 5\%ile & -0.35 & -0.47 & 0 & 0.2 & $\gamma_\rho = 2.79$ & $\lambda_\rho = 3.00$\\
         & 95\%ile & 0.93 & 0.17 & 3.85 & 2.05 & $\gamma_\eta = 0.70$ & $\lambda_\eta = 2.88$\\
         Table \ref{tab:sdr} & & &  & & & $\gamma_\rho = 1.07$ & $\lambda_\rho = 0.03$\\
         & & & & & & $\gamma_\eta = 1.40$ & $\lambda_\eta = 0.09$\\ 
         Hofstede & 5\%ile & 14.3 & 29.1 & 0 & 0.2 &  $\gamma_\rho = 4.78$ & $\lambda_\rho = 0.055$\\
          & 95\%ile & 87.0 & 95.9 & 4 & 4 &  $\gamma_\eta = -1.02$ & $\lambda_\eta = 0.042$\\ \hline
    \end{tabular}
    \caption*{Falk reports patience and risk-taking whereas Drupp reports impatience and risk aversion. We therefore flipped Falk's sign.}
\end{table}

\begin{table}
    \centering
     \caption{National time and risk preferences for four alternative calibrations.}
    \label{tab:assumptions}
    \tiny
    \begin{tabular}{l r r r r r r r r r r}
& \multicolumn{6}{c}{Falk} & \multicolumn{4}{c}{Table \ref{tab:sdr}} \\
& \multicolumn{2}{c}{unweighted} & \multicolumn{2}{c}{weighted} & \multicolumn{2}{c}{Eur \& NAm} & \multicolumn{2}{c}{observed} & \multicolumn{2}{c}{imputed} \\
& $\rho$ & $\eta$ & $\rho$ & $\eta$ & $\rho$ & $\eta$ & $\rho$ & $\eta$ & $\rho$ & $\eta$ \\ \hline
Afghanistan	&	3.18	&	1.42	&	4.20	&	2.18	&	3.40	&	0.35	&	1.07	&	1.39	&	1.07	&	1.39	\\
Algeria	&	2.24	&	0.65	&	3.01	&	0.46	&	2.61	&	0.00	&	1.06	&	1.37	&	1.06	&	1.37	\\
Argentina	&	3.28	&	1.65	&	4.33	&	2.69	&	3.48	&	0.58	&	1.11	&	1.32	&	1.07	&	1.40	\\
Australia	&	0.09	&	1.38	&	0.28	&	2.08	&	0.82	&	0.30	&	1.50	&	1.55	&	1.05	&	1.39	\\
Austria	&	0.27	&	1.95	&	0.50	&	3.34	&	0.97	&	0.88	&	1.00	&	1.60	&	1.05	&	1.41	\\
Bangladesh	&	2.16	&	2.34	&	2.91	&	4.21	&	2.55	&	1.27	&	1.06	&	1.42	&	1.06	&	1.42	\\
Bolivia	&	2.20	&	1.47	&	2.96	&	2.29	&	2.58	&	0.40	&	1.12	&	1.16	&	1.06	&	1.39	\\
Bosnia Herzegovina	&	3.35	&	2.13	&	4.41	&	3.75	&	3.53	&	1.06	&	1.07	&	1.41	&	1.07	&	1.41	\\
Botswana	&	1.61	&	0.00	&	2.21	&	0.00	&	2.09	&	0.00	&	1.06	&	1.34	&	1.06	&	1.34	\\
Brazil	&	3.39	&	2.49	&	4.47	&	4.54	&	3.57	&	1.42	&	1.11	&	2.72	&	1.07	&	1.42	\\
Cambodia	&	2.89	&	2.93	&	3.83	&	5.52	&	3.15	&	1.87	&	1.07	&	1.44	&	1.07	&	1.44	\\
Cameroon	&	4.00	&	3.31	&	5.24	&	6.34	&	4.07	&	2.24	&	1.08	&	1.45	&	1.08	&	1.45	\\
Canada	&	0.00	&	1.24	&	0.00	&	1.78	&	0.64	&	0.17	&	1.00	&	1.50	&	1.05	&	1.38	\\
Chile	&	3.02	&	1.41	&	3.99	&	2.15	&	3.26	&	0.34	&	1.12	&	1.28	&	1.07	&	1.39	\\
China	&	1.02	&	1.83	&	1.46	&	3.07	&	1.60	&	0.76	&	0.67	&	1.07	&	1.06	&	1.40	\\
Colombia	&	3.70	&	1.90	&	4.86	&	3.24	&	3.83	&	0.83	&	1.15	&	1.82	&	1.08	&	1.41	\\
Costa Rica	&	3.04	&	1.77	&	4.03	&	2.94	&	3.28	&	0.70	&	1.00	&	1.11	&	1.07	&	1.40	\\
Croatia	&	2.79	&	1.57	&	3.71	&	2.51	&	3.07	&	0.50	&	1.07	&	1.40	&	1.07	&	1.40	\\
Czech Republic	&	1.07	&	1.83	&	1.53	&	3.08	&	1.64	&	0.76	&	1.08	&	1.10	&	1.06	&	1.40	\\
Egypt	&	3.84	&	2.58	&	5.03	&	4.73	&	3.94	&	1.51	&	1.08	&	1.43	&	1.08	&	1.43	\\
Estonia	&	2.37	&	2.62	&	3.17	&	4.82	&	2.72	&	1.55	&	1.27	&	0.83	&	1.07	&	1.43	\\
Finland	&	0.30	&	2.58	&	0.54	&	4.74	&	0.99	&	1.52	&	1.00	&	1.60	&	1.05	&	1.43	\\
France	&	1.17	&	1.86	&	1.65	&	3.14	&	1.72	&	0.79	&	1.05	&	1.30	&	1.06	&	1.40	\\
Georgia	&	4.21	&	2.00	&	5.50	&	3.46	&	4.25	&	0.93	&	1.08	&	1.41	&	1.08	&	1.41	\\
Germany	&	0.21	&	1.90	&	0.43	&	3.23	&	0.92	&	0.83	&	0.75	&	1.50	&	1.05	&	1.41	\\
Ghana	&	2.15	&	0.00	&	2.90	&	0.00	&	2.54	&	0.00	&	1.06	&	1.35	&	1.06	&	1.35	\\
Greece	&	3.75	&	2.22	&	4.93	&	3.95	&	3.87	&	1.15	&	1.00	&	1.60	&	1.08	&	1.42	\\
Guatemala	&	3.38	&	2.40	&	4.46	&	4.34	&	3.56	&	1.33	&	1.00	&	1.04	&	1.07	&	1.42	\\
Haiti	&	3.81	&	1.72	&	5.00	&	2.83	&	3.92	&	0.64	&	1.08	&	1.40	&	1.08	&	1.40	\\
Hungary	&	4.01	&	3.20	&	5.25	&	6.11	&	4.08	&	2.14	&	1.30	&	1.10	&	1.08	&	1.45	\\
India	&	2.85	&	2.56	&	3.78	&	4.69	&	3.12	&	1.49	&	1.30	&	1.64	&	1.07	&	1.43	\\
Indonesia	&	3.76	&	2.69	&	4.94	&	4.99	&	3.88	&	1.63	&	1.08	&	1.43	&	1.08	&	1.43	\\
Iran	&	3.83	&	0.80	&	5.02	&	0.81	&	3.93	&	0.00	&	0.53	&	4.27	&	1.08	&	1.37	\\
Iraq	&	3.96	&	1.29	&	5.19	&	1.90	&	4.04	&	0.22	&	1.08	&	1.39	&	1.08	&	1.39	\\
Israel	&	0.81	&	1.07	&	1.20	&	1.40	&	1.42	&	0.00	&	1.05	&	1.38	&	1.05	&	1.38	\\
Italy	&	2.07	&	2.04	&	2.79	&	3.54	&	2.47	&	0.97	&	1.00	&	1.38	&	1.06	&	1.41	\\
Japan	&	2.07	&	2.79	&	2.79	&	5.21	&	2.47	&	1.73	&	1.50	&	1.40	&	1.06	&	1.43	\\
Jordan	&	3.96	&	2.13	&	5.20	&	3.74	&	4.05	&	1.06	&	1.08	&	1.41	&	1.08	&	1.41	\\
Kazakhstan	&	3.39	&	1.07	&	4.47	&	1.39	&	3.57	&	0.00	&	1.07	&	1.38	&	1.07	&	1.38	\\
Kenya	&	2.73	&	1.07	&	3.63	&	1.40	&	3.02	&	0.00	&	1.07	&	1.38	&	1.07	&	1.38	\\
Lithuania	&	2.68	&	1.90	&	3.57	&	3.24	&	2.98	&	0.83	&	1.37	&	0.53	&	1.07	&	1.41	\\
Malawi	&	2.62	&	0.41	&	3.49	&	0.00	&	2.93	&	0.00	&	1.07	&	1.36	&	1.07	&	1.36	\\
Mexico	&	2.85	&	2.17	&	3.78	&	3.83	&	3.12	&	1.10	&	1.00	&	2.71	&	1.07	&	1.41	\\
Moldova	&	1.75	&	1.87	&	2.39	&	3.17	&	2.21	&	0.80	&	1.06	&	1.41	&	1.06	&	1.41	\\
Morocco	&	3.58	&	1.97	&	4.70	&	3.39	&	3.72	&	0.90	&	1.07	&	1.41	&	1.07	&	1.41	\\
Netherlands	&	0.00	&	1.23	&	0.00	&	1.75	&	0.00	&	0.15	&	0.90	&	1.55	&	1.04	&	1.38	\\
Nicaragua	&	4.66	&	3.34	&	6.08	&	6.42	&	4.63	&	2.28	&	1.00	&	1.14	&	1.08	&	1.45	\\
Nigeria	&	3.18	&	0.66	&	4.20	&	0.50	&	3.39	&	0.00	&	1.07	&	1.37	&	1.07	&	1.37	\\
Pakistan	&	2.76	&	1.71	&	3.66	&	2.82	&	3.04	&	0.64	&	1.07	&	1.40	&	1.07	&	1.40	\\
Peru	&	2.85	&	1.33	&	3.78	&	1.97	&	3.12	&	0.25	&	1.13	&	1.39	&	1.07	&	1.39	\\
Philippines	&	2.10	&	0.92	&	2.83	&	1.08	&	2.49	&	0.00	&	1.06	&	1.37	&	1.06	&	1.37	\\
Poland	&	2.20	&	1.98	&	2.96	&	3.42	&	2.58	&	0.91	&	0.98	&	0.95	&	1.06	&	1.41	\\
Portugal	&	3.58	&	4.05	&	4.71	&	7.98	&	3.73	&	2.99	&	1.00	&	1.65	&	1.07	&	1.47	\\
Romania	&	3.42	&	2.43	&	4.51	&	4.40	&	3.60	&	1.36	&	1.07	&	1.42	&	1.07	&	1.42	\\
Russia	&	2.73	&	2.70	&	3.63	&	5.00	&	3.02	&	1.63	&	1.48	&	0.20	&	1.07	&	1.43	\\
Rwanda	&	4.64	&	2.60	&	6.05	&	4.79	&	4.61	&	1.54	&	1.08	&	1.43	&	1.08	&	1.43	\\
Saudi Arabia	&	1.74	&	0.00	&	2.37	&	0.00	&	2.19	&	0.00	&	1.06	&	1.34	&	1.06	&	1.34	\\
Serbia	&	2.95	&	2.14	&	3.91	&	3.77	&	3.21	&	1.07	&	1.07	&	1.41	&	1.07	&	1.41	\\
South Africa	&	2.25	&	0.00	&	3.02	&	0.00	&	2.62	&	0.00	&	1.06	&	1.31	&	1.06	&	1.31	\\
South Korea	&	1.13	&	1.88	&	1.60	&	3.20	&	1.68	&	0.81	&	1.10	&	1.00	&	1.06	&	1.41	\\
Spain	&	1.74	&	2.23	&	2.38	&	3.95	&	2.20	&	1.16	&	1.00	&	1.60	&	1.06	&	1.42	\\
Sri Lanka	&	2.82	&	1.59	&	3.74	&	2.55	&	3.09	&	0.52	&	1.07	&	1.40	&	1.07	&	1.40	\\
Suriname	&	2.43	&	1.26	&	3.25	&	1.82	&	2.77	&	0.19	&	1.07	&	1.39	&	1.07	&	1.39	\\
Sweden	&	0.00	&	1.62	&	0.00	&	2.62	&	0.00	&	0.55	&	1.10	&	1.25	&	1.04	&	1.40	\\
Switzerland	&	0.04	&	1.83	&	0.22	&	3.07	&	0.78	&	0.76	&	1.05	&	1.40	&	1.05	&	1.40	\\
Tanzania	&	3.63	&	0.36	&	4.77	&	0.00	&	3.77	&	0.00	&	1.08	&	1.36	&	1.08	&	1.36	\\
Thailand	&	3.28	&	2.13	&	4.33	&	3.73	&	3.48	&	1.06	&	1.07	&	1.41	&	1.07	&	1.41	\\
Turkey	&	2.63	&	1.70	&	3.50	&	2.80	&	2.93	&	0.63	&	0.60	&	1.32	&	1.07	&	1.40	\\
Uganda	&	3.38	&	1.30	&	4.45	&	1.92	&	3.56	&	0.23	&	1.07	&	1.39	&	1.07	&	1.39	\\
Ukraine	&	3.11	&	2.40	&	4.11	&	4.34	&	3.34	&	1.33	&	1.07	&	1.42	&	1.07	&	1.42	\\
United Arab Emirates	&	2.79	&	1.52	&	3.70	&	2.40	&	3.07	&	0.45	&	1.07	&	1.39	&	1.07	&	1.39	\\
United Kingdom	&	0.53	&	1.63	&	0.84	&	2.64	&	1.19	&	0.56	&	1.25	&	1.53	&	1.05	&	1.40	\\
United States	&	0.00	&	1.44	&	0.00	&	2.21	&	0.36	&	0.36	&	0.90	&	1.43	&	1.04	&	1.39	\\
Venezuela	&	3.28	&	1.05	&	4.32	&	1.37	&	3.47	&	0.00	&	1.00	&	1.71	&	1.07	&	1.38	\\
Vietnam	&	2.06	&	1.80	&	2.78	&	3.00	&	2.46	&	0.73	&	1.06	&	1.40	&	1.06	&	1.40	\\
Zimbabwe	&	3.32	&	0.27	&	4.37	&	0.00	&	3.51	&	0.00	&	1.07	&	1.35	&	1.07	&	1.35	\\ \\
average	&	2.51	&	1.75	&	3.34	&	2.97	&	2.81	&	0.79	&	1.07	&	1.42	&	1.07	&	1.40	\\
weighted	&	2.24	&	1.95	&	2.99	&	3.37	&	2.58	&	0.94	&	1.03	&	1.46	&	1.06	&	1.41	\\
 \hline
    \end{tabular}
\end{table}

\section{Additional results}
Table \ref{tab:results} shows the estimates of the global social cost of carbon for the SSP2 scenario, according to the preferences of each of the 76 countries. Four alternative calibrations of the national preferences are used; see Table \ref{tab:assumptions}. 

Table \ref{tab:aggresults} shows the aggregate results. The top row takes the average of the social cost of carbon\textemdash one country, one vote. The second row takes the population-weighted average\textemdash one person, one vote. The third row takes the average time and risk preferences instead, and the bottom row the population-weighted averages. The six columns show the six alternative calibrations. The first column calibrates to the Falk data, the second one to the population-weighted Falk data, the third one restricts the calibration to data from Europe and North America. The fourth column uses the observed preferences of Table \ref{tab:sdr} where available and imputed preferences elsewhere. The fifth column imputes all preferences. The sixth column uses the Hofstede data.

\begin{figure}
    \centering
    \caption{The social cost of carbon plotted against the calibrated pure rate of time preference.}
    \includegraphics[width=\textwidth]{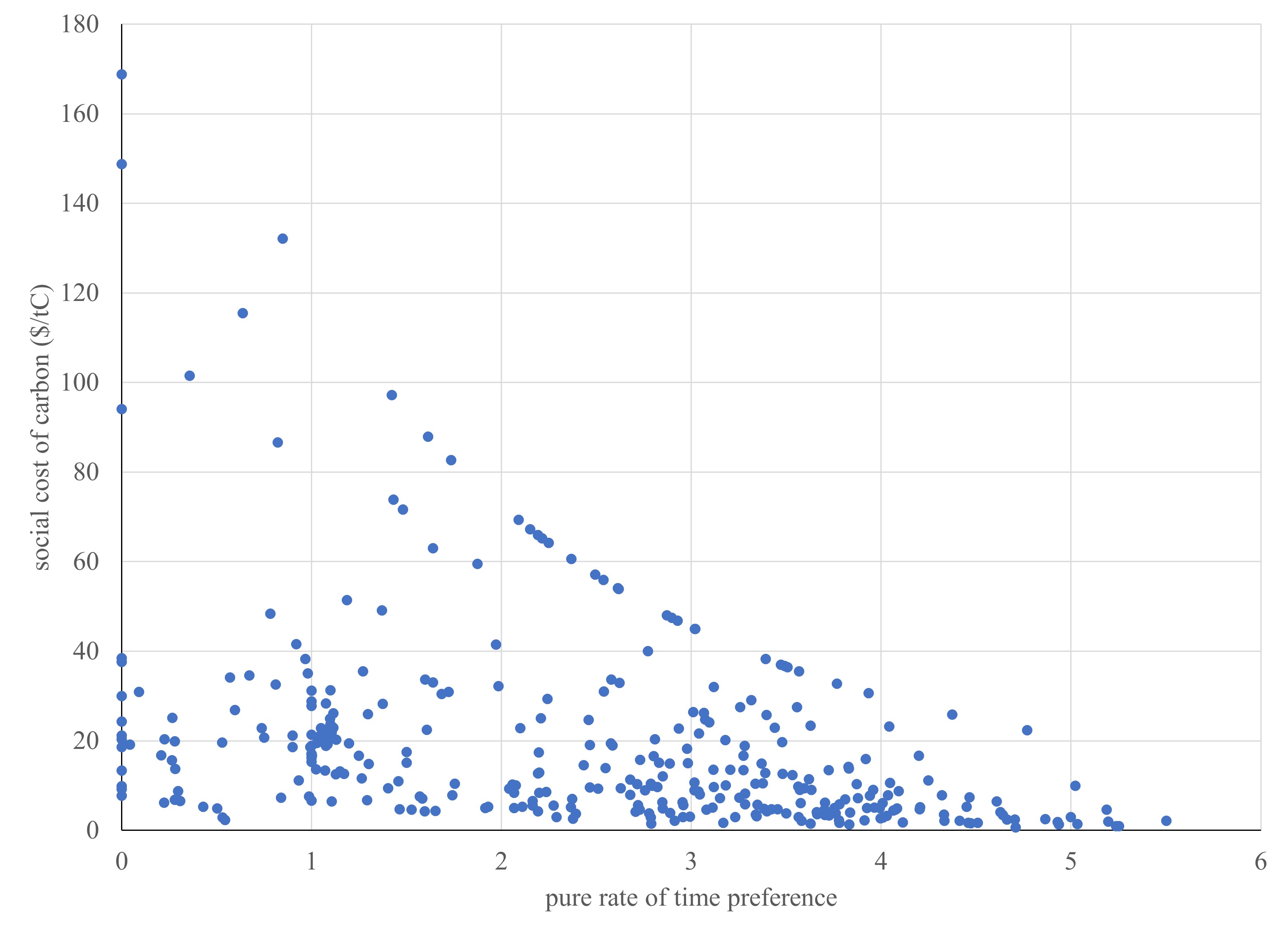}
    \label{fig:time}
\end{figure}

\begin{figure}
    \centering
    \caption{The social cost of carbon plotted against the calibrated inverse of the intertemporal rate of substitution.}
    \includegraphics[width=\textwidth]{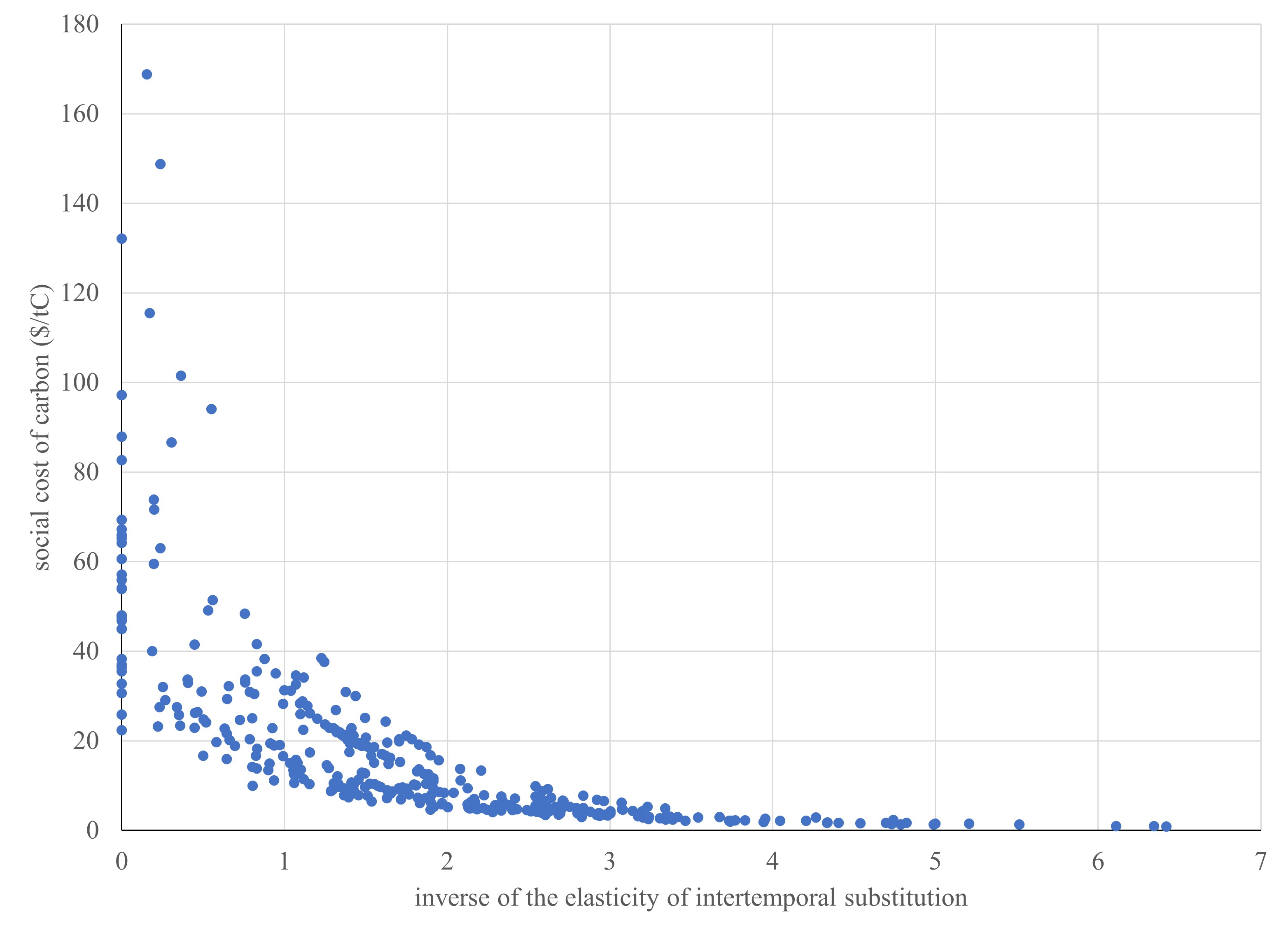}
    \label{fig:risk}
\end{figure}

In all six calibrations, the average social cost of carbon is larger than the social cost of carbon with average preferences. In other words, the social cost of carbon is a \emph{convex} function of time and risk preferences, at least in the domain considered here. See Figures \ref{fig:time} and \ref{fig:risk}.

\begin{table}
    \centering
    \caption{Alternative estimates of the social cost of carbon.}
    \label{tab:aggresults}
    \footnotesize
    \begin{tabular}{l c c c c c c} 
Calibration & \multicolumn{3}{c}{Falk} & \multicolumn{2}{c}{Table \ref{tab:sdr}} & Hofstede\\
& unweighted & weighted & NAM \& Eur & observed & imputed & \\ \hline
Average SCC	&	15.4	&	8.8	& 31.6 & 21.8 & 20.4 &	19.2\\
Weighted average SCC	&	12.2 & 27.9	&	5.9	& 22.6 &	20.3 & 51.4	\\
SCC for average preferences	&	9.3	&	3.5 & 20.3	& 20.0 &	20.4 & 10.0	\\
SCC for weighted preferences	&	8.6	&	3.1 & 18.9	& 19.5 &	20.3 & 22.5	\\ \hline
    \end{tabular}
\caption*{\footnotesize The social cost of carbon is for emissions in 2015, given in \$\textsuperscript{2015} per tonne of carbon. The top two rows are the (weighted) averages of the global social costs of carbon for national time and risk preferences; see Table \ref{tab:results}. The bottom two rows are for (weighted) average time and risk preference. Averaging is done with and without population weighting. The results are for three alternative calibrations of national time and risk preferences.}
\end{table}

\begin{figure}
    \centering
    \caption{The social cost of carbon for preferences calibrated to the Hofstede data versus the Falk data}
    \includegraphics[width=\textwidth]{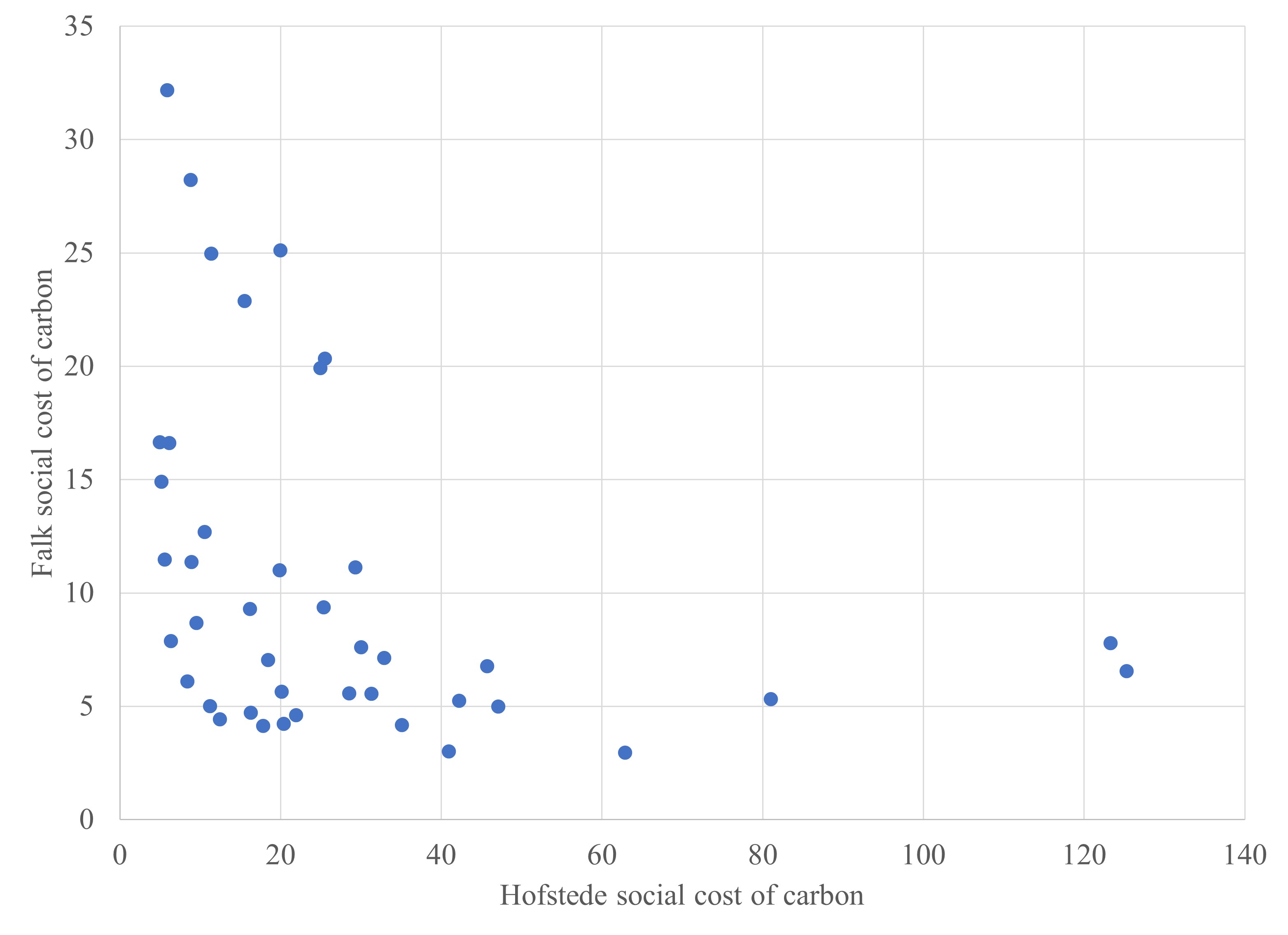}
    \label{fig:hofstede}
\end{figure}

The results for the Hofstede calibration are in Table \ref{tab:hofstede}. Figure \ref{fig:hofstede} plots the social cost of carbon against the unweighted Falk results for those countries that are in both datasets. The difference is clear. Hofstede's long-term orientation correlates with Falk's patience and uncertainty avoidance negatively correlates with risk-taking but those correlations are far from perfect: 0.29 and -0.33, respectively. However, as Figure \ref{fig:hofstede} shows, the correlation between the social cost of carbon when preferences are calibrated to Hofstede are negatively correlated (-0.28) to the Falk social cost of carbon. The bottom rows of Table \ref{tab:hofstede} corroborate the key qualitative result above: The social cost of carbon is very different when calibrated to the preferences observed in the rest of the world than when following people from Europe and North America.

The social cost of carbon is larger if we calibrate the Falk data to the risk and time preferences from the literature in Table \ref{tab:sdr}, primarily because the pure rate of time preference is much lower in this case, and particularly so in China.

The social cost of carbon, averaged over Drupp's experts, is higher than for all calibrations to the Falk data and the literature, and to three of the four Hofstede results. The exception is the population-weighted social cost of carbon for the Hofstede calibration. This is because people from China and India who work for foreign-owned companies are extraordinarily patient and risk-tolerant.

Table \ref{tab:scenarios} shows the social cost of carbon for alternative scenarios. The social cost of carbon is lowest for the SSP1 (SSP5) scenario for the Falk (literature) calibrations and highest for the SSP3 one. Our default scenario, SSP2, is in the middle of the range, regardless of calibration. The range in estimates is limited because scenarios deviate most from one another in the long run and these differences are discounted away. Table \ref{tab:impact} shows the sensitivity of the social cost of carbon to the impact function assumed. The base calibration is due to \citet{Barrage2023}. If instead we fit the same function to the meta-analysis in \citet{Tol2024data}, the social cost of carbon falls. Relying on the earlier meta-analysis of \citet{Howard2017} raises the social cost of carbon. In the base case, vulnerability to climate change falls with economic growth, assuming an income elasticity of $\varepsilon=-0.36$. More (less) elastic vulnerability lowers (raises) the social cost of carbon. If we set the income elasticity to zero, the social cost of carbon goes up further. If vulnerability \emph{increases} with development, as assumed by \citet{Sterner2008, Drupp2021} and \citet{Bremer2021}, the social cost of carbon is higher still \$30.9/tC.

Table \ref{tab:cluster} uses k-means clustering on risk and time to group the 76 countries into 4 clusters. Cluster 3 combines a high time preference with a low risk aversion, cluster 4 is the opposite. Cluster 3 is the poorest, cluster 4 the richest.\footnote{Income was not used to cluster. Note that China clusters with the rich countries.} Cluster 2 has high values for both, cluster 1 lower values. Incomes are about the same, in between clusters 3 and 4.\footnote{Three-way clustering would allocate the countries in cluster 3 to clusters 1 and 2. Five-way clustering would split cluster 1. Four-way clustering leads to identical results with and without population-weighting of preferences.} Cluster 2, which includes India, advocates the lowest social cost of carbon, in line with the high rates of time preference and risk aversion. Cluster 1 is the second-lowest. Cluster 3, the poorest cluster, favours the highest social cost of carbon: Their low rate of risk aversion dominates their high time preference. Cluster 4 is in between. For all four clusters, population weighting reduces the average social cost of carbon.  

Clustering has little effect for the calibrations to the literature summarized in Table \ref{tab:sdr} because there is so little variation between countries.

\begin{table}
    \centering
    \caption{Welfare parameters and the implied social cost of carbon for four clusters of countries.}
    \begin{tabular}{l c c c c}
cluster & 1 & 2 & 3 & 4 \\ \hline
\multicolumn{5}{c}{Falk \& Drupp} \\
Social cost of carbon (\$/tC) & 9.76 & 4.55 & 45.98 & 21.72 \\
pure time preference & 2.73 & 3.63 & 2.48 & 0.44 \\
risk aversion & 1.69 & 2.54 & 0.33 & 1.68 \\
\multicolumn{5}{c}{Falk \& Drupp, population-weighted} \\
Social cost of carbon (\$/tC) & 4.00 & 1.63 & 25.97 & 6.72 \\
pure time preference & 3.56 & 4.14 & 3.58 & 1.12 \\
risk aversion& 3.04 & 4.67 & 0.42 & 2.89 \\
\multicolumn{5}{c}{Falk \& Drupp, Europe \& North America} \\
Social cost of carbon (\$/tC) & 21.52 & 9.36 & 47.00 & 48.62 \\
pure time preference & 2.97 & 3.36 & 2.99 & 1.32 \\
risk aversion& 0.76 & 1.48 & 0.00 & 0.67 \\
\multicolumn{5}{c}{observed and imputed} \\
Social cost of carbon (\$/tC) & 21.87 & 21.96 & 21.51 & 21.83 \\
pure time preference & 1.06 & 1.11 & 1.07 & 1.03 \\
risk aversion& 1.46 & 1.43 & 1.35 & 1.38 \\
\multicolumn{5}{c}{imputed} \\
Social cost of carbon (\$/tC) & 20.43 & 19.81 & 21.51 & 20.57 \\
pure time preference & 1.07 & 1.08 & 1.07 & 1.05 \\
risk aversion& 1.40 & 1.43 & 1.35 & 1.40 \\ \\
income (US\$/person/year) &	16,502 & 17,034 & 3,553 & 42,583 \\
income (Geary-Khamis \$/person/year) & 9,438 & 6,257 & 1,529 & 34,997 \\ 
population (millions) & 1,449 & 2,495 & 590 & 2,147 \\\hline
    \end{tabular}
    \caption*{\footnotesize The 2015 social cost of carbon  (\$\textsuperscript{2015}/tC) is the average of the global social costs of carbon for national time and risk preferences in the respective clusters. Averages are unweighted unless indicated otherwise. Clusters are found by k-means clustering on time and risk preferences.}
    \label{tab:cluster}
\end{table}

\begin{table}
    \centering
     \caption{The 2015 global social cost of carbon (\$\textsuperscript{2015}) per tonne of carbon) for national time and risk preferences according to four alternative calibrations.}
    \label{tab:results}
    \tiny
    \begin{tabular}{l r r r r r}
& \multicolumn{3}{c}{Falk}	& \multicolumn{2}{c}{Table \ref{tab:sdr}}\\
& unweighted & weighted  & Eur \& NAm & observed & imputed \\ \hline
Afghanistan	&	10.1	&	4.7	&	25.7	&	20.6	&	20.6	\\
Algeria	&	29.3	&	26.4	&	54.0	&	21.2	&	21.2	\\
Argentina	&	8.2	&	3.5	&	19.7	&	21.9	&	20.4	\\
Australia	&	30.9	&	13.7	&	86.6	&	15.1	&	20.8	\\
Austria	&	15.6	&	4.9	&	38.3	&	17.1	&	20.4	\\
Bangladesh	&	6.5	&	2.2	&	13.9	&	20.0	&	20.0	\\
Bolivia	&	12.9	&	5.6	&	33.7	&	26.1	&	20.6	\\
Bosnia Herzegovina	&	5.8	&	2.1	&	12.3	&	20.1	&	20.1	\\
Botswana	&	87.9	&	65.2	&	69.3	&	21.9	&	21.9	\\
Brazil	&	4.6	&	1.6	&	9.1	&	6.4	&	19.9	\\
Cambodia	&	3.9	&	1.3	&	7.2	&	19.6	&	19.6	\\
Cameroon	&	2.7	&	0.9	&	4.6	&	19.3	&	19.3	\\
Canada	&	37.7	&	20.3	&	115.5	&	18.9	&	20.9	\\
Chile	&	10.7	&	5.0	&	27.5	&	22.9	&	20.6	\\
China	&	13.6	&	4.7	&	33.7	&	34.6	&	20.4	\\
Colombia	&	6.2	&	2.5	&	13.8	&	13.2	&	20.2	\\
Costa Rica	&	8.0	&	3.2	&	18.8	&	28.8	&	20.4	\\
Croatia	&	9.9	&	4.2	&	24.8	&	20.5	&	20.5	\\
Czech Republic	&	13.4	&	4.6	&	33.0	&	28.3	&	20.4	\\
Egypt	&	4.0	&	1.4	&	7.7	&	19.8	&	19.8	\\
Estonia	&	5.2	&	1.7	&	10.3	&	35.5	&	19.8	\\
Finland	&	8.8	&	2.3	&	18.6	&	17.1	&	20.0	\\
France	&	12.6	&	4.4	&	30.9	&	22.9	&	20.4	\\
Georgia	&	5.2	&	2.1	&	11.2	&	20.1	&	20.1	\\
Germany	&	16.8	&	5.3	&	41.5	&	20.7	&	20.4	\\
Ghana	&	67.2	&	47.5	&	55.9	&	21.7	&	21.7	\\
Greece	&	5.0	&	1.9	&	10.4	&	17.1	&	20.0	\\
Guatemala	&	4.8	&	1.7	&	9.8	&	31.2	&	19.9	\\
Haiti	&	6.9	&	2.9	&	15.9	&	20.3	&	20.3	\\
Hungary	&	2.8	&	1.0	&	4.9	&	25.9	&	19.4	\\
India	&	4.9	&	1.6	&	9.7	&	14.8	&	19.8	\\
Indonesia	&	3.8	&	1.3	&	7.2	&	19.7	&	19.7	\\
Iran	&	14.2	&	10.0	&	30.6	&	2.9	&	21.0	\\
Iraq	&	9.0	&	4.6	&	23.2	&	20.6	&	20.6	\\
Israel	&	32.6	&	19.4	&	97.1	&	21.0	&	21.0	\\
Italy	&	8.4	&	2.9	&	19.1	&	21.3	&	20.2	\\
Japan	&	5.0	&	1.5	&	9.6	&	17.5	&	19.7	\\
Jordan	&	5.1	&	2.0	&	10.6	&	20.1	&	20.1	\\
Kazakhstan	&	12.8	&	7.4	&	35.5	&	20.8	&	20.8	\\
Kenya	&	15.8	&	9.1	&	44.9	&	20.9	&	20.9	\\
Lithuania	&	7.9	&	3.0	&	18.3	&	49.1	&	20.3	\\
Malawi	&	33.0	&	36.7	&	46.8	&	21.3	&	21.3	\\
Mexico	&	6.3	&	2.2	&	13.5	&	6.6	&	20.1	\\
Moldova	&	10.4	&	3.7	&	25.1	&	20.4	&	20.4	\\
Morocco	&	6.1	&	2.4	&	13.4	&	20.2	&	20.2	\\
Netherlands	&	38.4	&	21.2	&	168.8	&	18.6	&	21.0	\\
Nicaragua	&	2.4	&	0.9	&	4.1	&	27.8	&	19.2	\\
Nigeria	&	20.2	&	16.7	&	38.2	&	21.1	&	21.1	\\
Pakistan	&	9.0	&	3.6	&	21.6	&	20.4	&	20.4	\\
Peru	&	12.0	&	5.8	&	32.0	&	20.3	&	20.7	\\
Philippines	&	22.9	&	15.1	&	57.1	&	21.0	&	21.0	\\
Poland	&	8.4	&	3.0	&	19.4	&	35.1	&	20.3	\\
Portugal	&	2.1	&	0.7	&	3.3	&	16.2	&	18.9	\\
Romania	&	4.7	&	1.7	&	9.4	&	19.9	&	19.9	\\
Russia	&	4.6	&	1.5	&	8.9	&	71.6	&	19.8	\\
Rwanda	&	3.5	&	1.3	&	6.5	&	19.7	&	19.7	\\
Saudi Arabia	&	82.6	&	60.6	&	66.0	&	21.8	&	21.8	\\
Serbia	&	6.2	&	2.3	&	13.5	&	20.1	&	20.1	\\
South Africa	&	64.2	&	45.0	&	53.9	&	22.4	&	22.4	\\
South Korea	&	12.5	&	4.3	&	30.4	&	31.3	&	20.4	\\
Spain	&	7.9	&	2.6	&	17.4	&	17.1	&	20.1	\\
Sri Lanka	&	9.7	&	4.1	&	24.1	&	20.5	&	20.5	\\
Suriname	&	14.6	&	7.3	&	40.0	&	20.7	&	20.7	\\
Sweden	&	24.3	&	9.3	&	94.1	&	23.7	&	20.7	\\
Switzerland	&	19.1	&	6.2	&	48.4	&	20.5	&	20.5	\\
Tanzania	&	23.4	&	22.3	&	32.8	&	21.3	&	21.3	\\
Thailand	&	5.9	&	2.2	&	12.6	&	20.1	&	20.1	\\
Turkey	&	9.4	&	3.8	&	22.7	&	26.9	&	20.4	\\
Uganda	&	10.5	&	5.2	&	27.5	&	20.7	&	20.7	\\
Ukraine	&	5.1	&	1.8	&	10.4	&	19.9	&	19.9	\\
United Arab Emirates	&	10.4	&	4.5	&	26.3	&	20.5	&	20.5	\\
United Kingdom	&	19.6	&	7.3	&	51.5	&	16.7	&	20.6	\\
United States	&	30.0	&	13.3	&	101.5	&	21.2	&	20.8	\\
Venezuela	&	13.4	&	7.8	&	37.0	&	15.3	&	20.8	\\
Vietnam	&	10.2	&	3.8	&	24.7	&	20.4	&	20.4	\\
Zimbabwe	&	29.1	&	25.8	&	36.4	&	21.4	&	21.4	\\ \hline
    \end{tabular}
\end{table}

\begin{table}
    \centering
    \caption{The global average social cost of carbon for alternative scenarios.}
    \label{tab:scenarios}
    \begin{tabular}{l c c c c c c}
    & \multicolumn{3}{c}{Falk} & \multicolumn{2}{c}{Table \ref{tab:sdr}} & Hofstede \\
    & unweighted & weighted & Eur \& NAm & observed & imputed & \\ \hline
    SSP1 &  14.5 & 5.4 & 34.2 & 18.3 & 16.5 & 18.9\\
    SSP2 &  15.4 & 5.9 & 31.6 & 21.8 & 20.4 & 19.2\\
    SSP3 & 18.1 & 8.2 & 27.1 & 29.2 & 28.7 & 20.8\\
    SSP4 & 15.1 & 5.9 & 28.9 & 22.6 & 21.4 & 18.3\\
    SSP5 & 15.4 & 5.5 & 40.3 & 16.6 & 14.3 & 21.0\\ \hline
    \end{tabular}
    \caption*{\footnotesize The 2015 social cost of carbon  (\$\textsuperscript{2015}/tC) is the average of the global social costs of carbon for national time and risk preferences. Averages are unweighted unless indicated otherwise.}
\end{table}

\begin{table}
    \centering
    \caption{The global average social cost of carbon for alternative impact functions.}
    \label{tab:impact}
    \footnotesize
    \begin{tabular}{l c c c c c c}
    & \multicolumn{3}{c}{Falk} & \multicolumn{2}{c}{Table \ref{tab:sdr}} & Hofstede \\
    & unweighted & weighted & Eur \& NAm & observed & imputed \\ \hline
    \citet{Tol2024EnPol} &  9.8 & 3.8 & 20.1 & 13.9 & 13.0 & 12.2\\
    \citet{Barrage2023} &  15.4 & 5.9 & 31.6 & 21.8 & 20.4 & 19.2\\
    \citet{Howard2017} & 37.9 & 14.6 & 78.0 & 53.9 & 50.4 & 47.3\\ \hline
    \end{tabular}
    \caption*{\footnotesize The 2015 social cost of carbon  (\$\textsuperscript{2015}/tC) is the average of the global social costs of carbon for national time and risk preferences. Averages are unweighted unless indicated otherwise.}
\end{table}

\begin{table}
    \centering
    \caption{The global average social cost of carbon for alternative income elasticities.}
    \label{tab:elasticities}
    \begin{tabular}{l c c c c c c}
    & \multicolumn{3}{c}{Falk} & \multicolumn{2}{c}{Table \ref{tab:sdr}} & Hofstede \\
    & unweighted & weighted & Eur \& NAm & observed & imputed \\ \hline
    $\varepsilon = 0.18$ &  30.9 & 10.6 & 68.7 & 44.5 & 40.9 & 40.4\\
    $\varepsilon = 0.00$ &  24.3 & 8.7 & 52.6 & 34.8 & 32.2 & 31.2\\
    $\varepsilon = -0.18$ &  19.2 & 7.1 & 40.6 & 27.4 & 25.5 & 24.4\\
    $\varepsilon = -0.36$ &  15.4 & 5.9 & 31.6 & 21.8 & 20.4 & 19.2\\
    $\varepsilon = -0.72$ &  10.1 & 4.2 & 19.8 & 14.2 & 13.4 & 12.3\\ \hline
    \end{tabular}
    \caption*{\footnotesize The 2015 social cost of carbon  (\$\textsuperscript{2015}/tC) is the average of the global social costs of carbon for national time and risk preferences. Averages are unweighted unless indicated otherwise.}
\end{table}

\begin{table}
    \tiny
    \centering
    \caption{Social cost of carbon when preferences are calibrated to Hofstede's cultural dimensions}
    \begin{tabular}{l c c c c c}
      &   		LTO	&	UA	&	PRTP &	RRA &	SCC	\\ \hline
Argentina	&	20.40	&	86.00	&	3.66	&	2.59	&	4.1	\\
Australia	&	21.16	&	51.00	&	3.62	&	1.12	&	11.5	\\
Austria	&	60.45	&	70.00	&	1.46	&	1.91	&	11.0	\\
Bangladesh	&	47.10	&	60.00	&	2.19	&	1.50	&	12.7	\\
Belgium	&	81.86	&	94.00	&	0.28	&	2.92	&	6.8	\\
Brazil	&	43.83	&	76.00	&	2.37	&	2.17	&	7.0	\\
Bulgaria	&	69.02	&	85.00	&	0.99	&	2.54	&	7.5	\\
Canada	&	36.02	&	48.00	&	2.80	&	0.99	&	16.6	\\
Chile	&	30.98	&	86.00	&	3.08	&	2.59	&	4.6	\\
China	&	87.41	&	30.00	&	0.00	&	0.24	&	148.7	\\
Colombia	&	13.10	&	80.00	&	4.06	&	2.33	&	4.4	\\
Croatia	&	58.44	&	80.00	&	1.57	&	2.33	&	7.6	\\
Czech Republic	&	70.03	&	74.00	&	0.93	&	2.08	&	11.1	\\
Denmark	&	34.76	&	23.00	&	2.87	&	0.00	&	48.0	\\
El Salvador	&	19.65	&	94.00	&	3.70	&	2.92	&	3.4	\\
Estonia	&	82.12	&	60.00	&	0.27	&	1.50	&	25.1	\\
Finland	&	38.29	&	59.00	&	2.68	&	1.45	&	11.4	\\
France	&	63.48	&	86.00	&	1.29	&	2.59	&	6.8	\\
Germany	&	82.87	&	65.00	&	0.22	&	1.70	&	20.3	\\
Great Britain	&	51.13	&	35.00	&	1.97	&	0.45	&	41.5	\\
Greece	&	45.34	&	112.00	&	2.29	&	3.67	&	2.9	\\
Hong Kong	&	60.96	&	29.00	&	1.43	&	0.20	&	73.8	\\
Hungary	&	58.19	&	82.00	&	1.58	&	2.42	&	7.1	\\
India	&	50.88	&	40.00	&	1.98	&	0.66	&	32.2	\\
Indonesia	&	61.96	&	48.00	&	1.37	&	0.99	&	28.2	\\
Iran	&	13.60	&	59.00	&	4.04	&	1.45	&	7.9	\\
Ireland	&	24.43	&	35.00	&	3.44	&	0.45	&	22.9	\\
Israel	&	37.53	&	81.00	&	2.72	&	2.38	&	5.6	\\
Italy	&	61.46	&	75.00	&	1.40	&	2.12	&	9.4	\\
Japan	&	87.91	&	92.00	&	0.00	&	2.84	&	7.8	\\
South Korea	&	100.00	&	85.00	&	0.00	&	2.54	&	9.9	\\
Latvia	&	68.77	&	63.00	&	1.00	&	1.62	&	16.7	\\
Lithuania	&	81.86	&	65.00	&	0.28	&	1.70	&	19.9	\\
Luxembourg	&	63.98	&	70.00	&	1.26	&	1.91	&	11.6	\\
Malaysia	&	40.81	&	36.00	&	2.54	&	0.49	&	31.0	\\
Malta	&	47.10	&	96.00	&	2.19	&	3.00	&	4.3	\\
Mexico	&	24.18	&	82.00	&	3.45	&	2.42	&	4.7	\\
Morocco	&	14.11	&	68.00	&	4.01	&	1.83	&	6.1	\\
Netherlands	&	67.00	&	53.00	&	1.10	&	1.20	&	25.0	\\
New Zealand	&	32.75	&	49.00	&	2.98	&	1.03	&	15.0	\\
Norway	&	34.51	&	50.00	&	2.89	&	1.08	&	14.9	\\
Pakistan	&	49.87	&	70.00	&	2.04	&	1.91	&	9.3	\\
Peru	&	25.19	&	87.00	&	3.40	&	2.63	&	4.2	\\
Philippines	&	27.46	&	44.00	&	3.27	&	0.82	&	16.6	\\
Poland	&	37.78	&	93.00	&	2.71	&	2.88	&	4.2	\\
Portugal	&	28.21	&	104.00	&	3.23	&	3.34	&	3.0	\\
Romania	&	51.89	&	90.00	&	1.93	&	2.75	&	5.2	\\
Russia	&	81.36	&	95.00	&	0.31	&	2.96	&	6.5	\\
Serbia	&	52.14	&	92.00	&	1.92	&	2.84	&	5.0	\\
Singapore	&	71.54	&	8.00	&	0.85	&	0.00	&	132.1	\\
Slovak Rep	&	76.57	&	51.00	&	0.57	&	1.12	&	34.1	\\
Slovenia	&	48.61	&	88.00	&	2.11	&	2.67	&	5.3	\\
Spain	&	47.61	&	86.00	&	2.16	&	2.59	&	5.5	\\
Sweden	&	52.90	&	29.00	&	1.87	&	0.20	&	59.5	\\
Switzerland	&	73.55	&	58.00	&	0.74	&	1.41	&	22.9	\\
Taiwan	&	92.95	&	69.00	&	0.00	&	1.87	&	18.6	\\
Thailand	&	31.74	&	64.00	&	3.04	&	1.66	&	8.7	\\
Trinidad and Tobago	&	12.59	&	55.00	&	4.09	&	1.29	&	8.8	\\
Turkey	&	45.59	&	85.00	&	2.28	&	2.54	&	5.6	\\
United States	&	25.69	&	46.00	&	3.37	&	0.91	&	14.9	\\
Uruguay	&	26.20	&	100.00	&	3.34	&	3.17	&	3.2	\\
Venezuela	&	15.62	&	76.00	&	3.93	&	2.17	&	5.0	\\
Vietnam	&	57.18	&	30.00	&	1.64	&	0.24	&	63.0	\\
											\\
Average	&	49.55	&	67.17	&	2.08	&	1.81	&	10.0	\\
Weighted average	&	58.02	&	50.95	&	1.61	&	1.12	&	22.5	\\
											\\
North America	&	26.75	&	46.21	&	3.31	&	0.92	&	15.1	\\
Europe	&	59.27	&	70.33	&	1.52	&	1.93	&	17.2	\\
Rest of the world	&	60.23	&	49.21	&	1.49	&	1.04	&	57.8	\\ \hline
    \end{tabular}
    \caption*{The table shows Hofstede's long-term orientation (LTO) and uncertainty avoidance (UA), the calibrated pure rate of time preference (PRTP) and rate of relative risk aversion (RRA), and the resulting estimate of the social cost of carbon (SCC; in dollar per tonne of carbon).}
    \label{tab:hofstede}
\end{table}

\end{document}